\documentstyle[11pt,dunk2001_asp,twoside]{article}
\def\edcomment#1{\iffalse\marginpar{\raggedright\sl#1\/}\else\relax\fi}
\reversemarginpar

\begin{document}
\title{Star Formation in the Local Universe}
 \author{Daniela Calzetti}
\affil{Space Telescope Science Institute, 3700 San Martin Drive, Baltimore, 
MD 21218, U.S.A., calzetti@stsci.edu}
\author{Jason Harris}
\affil{Space Telescope Science Institute, jharris@stsci.edu }

\begin{abstract}
We present preliminary results of a long-term study aimed at answering a
number of open questions on the evolution of starbursts in local galaxies. The
project employes mainly HST data from the ultraviolet to the red of the
stellar continuum and of the nebular emission from the galaxies. Here we
concentrate on NGC5253 and NGC5236 (M83), that form a dwarf--massive galaxy
pair at about 4~Mpc distance. The recent star formation history of the centers
of the two galaxies is investigated in order to identify similarities and
differences in the evolution of their central starbursts.
\end{abstract}

\section{Introduction}

Star formation is a driving force behind galaxy evolution; it largely
determines the stellar, energy (radiant and mechanical), metal, and
dust content of a galaxy.  Throughout much of the history of the
Universe, a significant fraction of the galaxy population is
represented by galaxies undergoing intense star formation activity; at
redshift z$\simeq$1 the star formation rate per unit comoving volume
was roughly an order of magnitude higher than at present times (Lilly
et al.\ 1996; Madau et al.\ 1996; Cowie, Songaila \& Barger 1999). Yet,
a number of questions related to the evolution of intense star
formation episodes remain unresolved. Among them: (1) the duration,
duty cycle, and star formation history of starbursts; (2) the
importance of self-triggering and propagation for the evolution of the
starburst (e.g., Walborn et al.\ 1999); (3) the impact of starbursts on
the host galaxy's ISM and the feedback onto the evolution of the
starburst itself; and (4) the possibility of distinct cluster and
diffuse field star formation modes (Meurer et al.\ 1995).

In order to begin addressing these issues, we are undertaking a comprehensive
study of $\sim$20 nearby ($<$20 Mpc) starburst galaxies covering a
representative range of characteristics: morphological types ranging from
dwarf irregulars (He~2-10) to grand-design spirals (M83); starburst star
formation rates (SFR) from 0.1~M$_{\odot}$~yr$^{-1}$ to
5~M$_{\odot}$~yr$^{-1}$; metallicities from 1/13~Z$_{\odot}$ to 2~Z$_{\odot}$;
and dust attenuations in the range 0$\la$A$_V\la$3. The data that are being
collected for this study include HST ultraviolet and optical imaging in
broad-- and narrow--band filters of the stellar continuum and selected nebular
emission lines (typically H$\beta$ and H$\alpha$ for dust reddening
corrections and for tracing the ionized gas, [OIII](5007~\AA) and
[SII](6731~\AA) for mapping of photo-- and shock--ionized gas), ground--based
optical imaging, and HST UV spectroscopy. In this contribution, we concentrate
on the galaxy pair NGC5253--NGC5236, for which almost all needed data are
available and the analysis is almost completed.

The two galaxies form an interesting, complementary pair in the Centaurus
Group: NGC5253 is a dwarf galaxy that is about 10 times more metal poor
($\sim$1/5~Z$_{\odot}$) and about 100 times less massive
($\approx$10$^9$~M$_{\odot}$) than its grand-design spiral companion NGC5236
(M83). The projected distance between the two galaxies is $\sim$130~kpc, and a
close encounter between the two about 1--2~Gyr ago has been suggested as the
initial trigger of the star formation in NGC5253 (Rogstad, Lockhart \& Wright
1974; van den Bergh 1980; Caldwell \& Phillips 1989). Despite the
dissimilarities of the host galaxies, the central starbursts share some common
properties: both have a SFR$\sim$0.2~M$_{\odot}$~yr$^{-1}$ over an area of
$\sim$400~pc in size; and both are characterized by patchy dust obscuration,
with relatively transparent regions counterbalanced by highly opaque areas and
dust lanes. However, while in NGC5253 star formation is diffused throughout
the central region (Calzetti et al.\ 1997, and references therein), in NGC5236
star formation is concentrated in a semi--annulus centered on the optical
nucleus, and has possibly been triggered by the galaxy's bar (Thatte et
al. 2000; Harris et al.\ 2001).

Located at a distance of about 4~Mpc (Sandage et al.\ 1994), the two
galaxies were imaged with the HST WFPC2 from the UV to the I-band
(Calzetti et al.\ 1997, Harris et al.\ 2001), yielding a resolution of
$\sim$1--2~pc per pixel, which is comparable to the typical half-light
radius of stellar clusters. The UV--V--I colors of the central
galaxies' regions, combined with the H$\alpha$ emission map, were used
to constrain the ages of the stellar populations, while
H$\alpha$/H$\beta$ ratio maps were used to remove the effects of dust
reddening from the colors and luminosities. HST STIS long-slit UV
spectra of the central region of NGC5253 were used to analyze in
detail the characteristics of the hot star populations in this galaxy
(Tremonti et al.\ 2001). Below is a summary of our current understanding 
of the star formation history in the centers of the two galaxies.
 
\section{The Dwarf Galaxy NGC5253}

\def\putplot#1#2#3#4#5#6#7{\begin{centering} \leavevmode
\vbox to#2{\rule{0pt}{#2}}
\includegraphics{#1}
\end{centering}}
%

The central region of active star formation in this galaxy is very
blue at UV--optical wavelengths, except for a prominent dust lane with
A$_V\ga$2.2~mag that bisects the starburst nearly perpendicularly to
the galaxy's major axis (Calzetti et al.\ 1997; and Figure~1). The region
contains about a dozen UV-bright stellar clusters, as well as
diffusely distributed blue stars (Meurer et al.\ 1995). Radio
observations, however, indicate that the bulk of the most recent star
formation is hidden by dust (Turner, Ho \& Beck 1998; Turner, Beck \&
Ho 2000). The distribution of the UV emission is slightly elongated
along the major axis, and is markedly different from the morphology of
the ionized gas emission (H$\alpha$); the latter is circularly
symmetric about a bright stellar cluster close to the geometric center
of the galaxy (NGC5253--5 of Calzetti et al.\ 1997, see Figure~1), and is
about a factor of 2 more extended than the UV stellar continuum
(Calzetti et al.\ 1999).

\begin{figure}[t]
\putplot{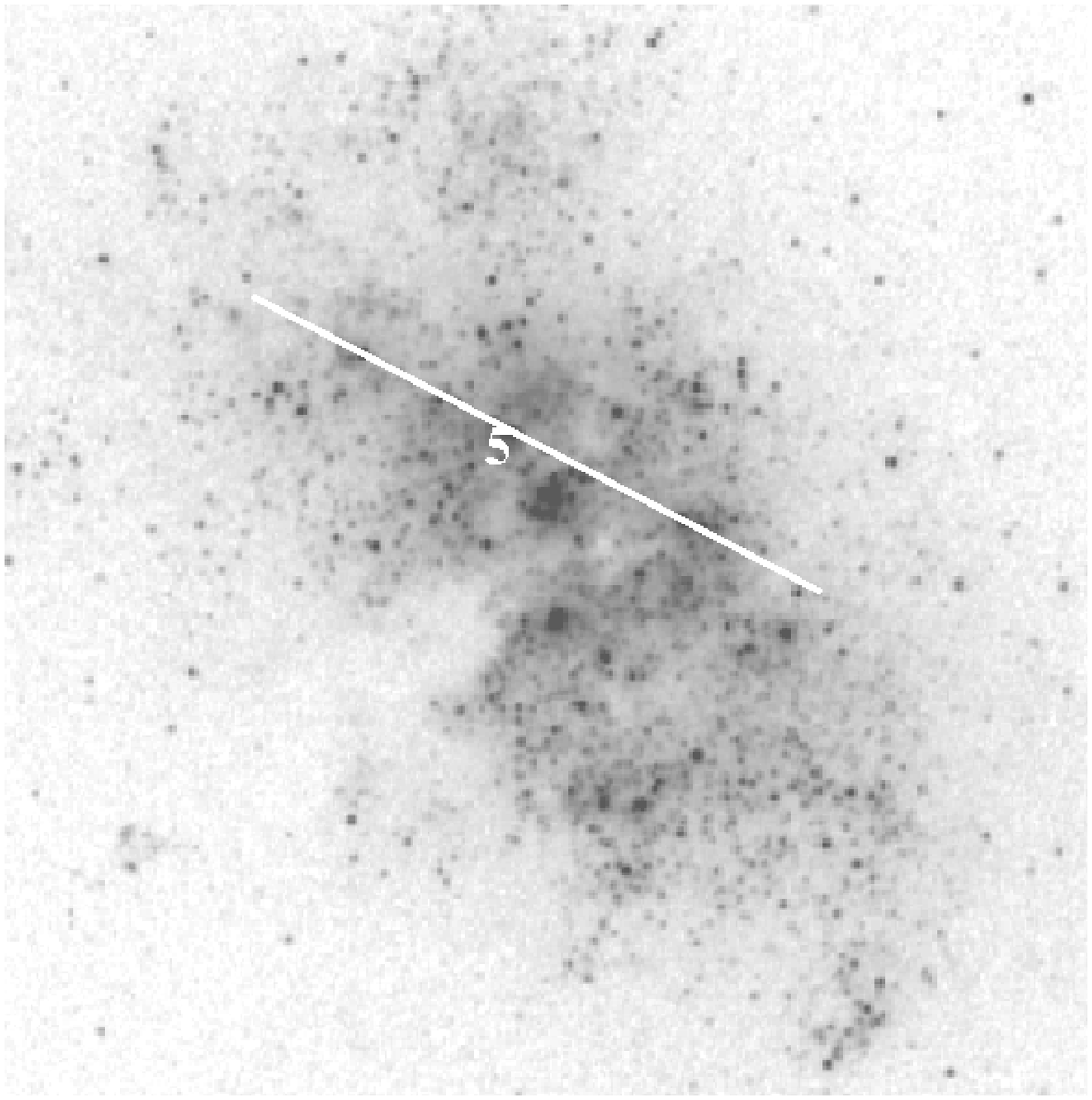}{4.0 cm}{0}{32}{32}{0}{-95}
\putplot{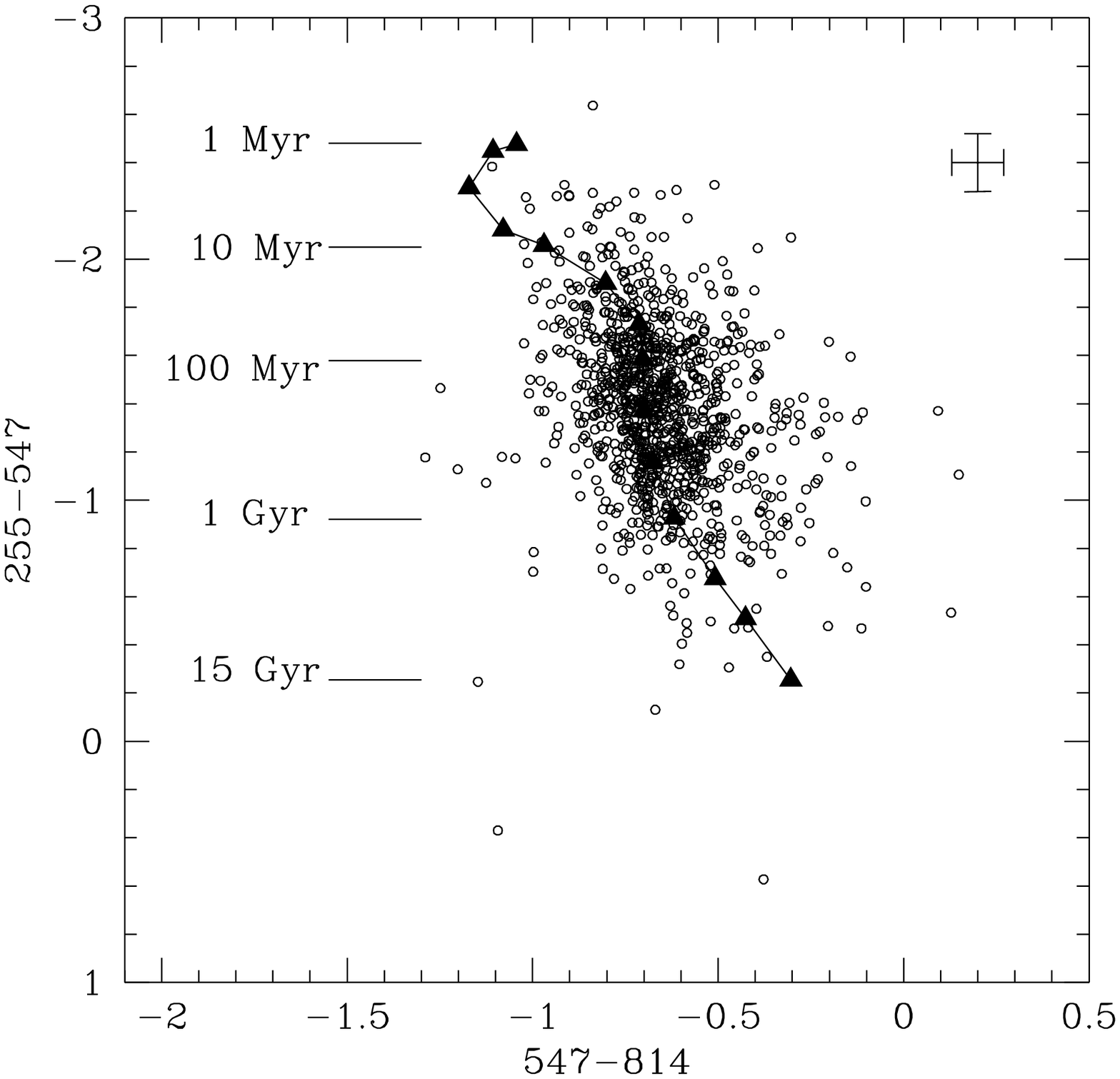}{4.0 cm}{0}{34}{34}{180}{10} 
{\bf Figure 1:} (left) The UV (2600~\AA) WFPC2 image of the central
starburst in NGC5253, showing the presence of a number of bright
stellar clusters and of diffuse emission. The image has size
38$^{\prime\prime}$, which correspond to a physical size of
$\sim$700~pc. The position of the STIS slit is shown as a thick line
(Tremonti et al.\ 2001). The number `5' shows the position of the
super-star-cluster candidate NGC5253-5, which has age $\la$2~Myr and
is at the center of the peak-activity region (Calzetti et
al. 1997). North is up, East is left.\\ 
\hspace*{1.4mm} {\bf Figure~2:} (right)
Color--color (UV$-$V versus V$-$I) diagram of the central 400~pc of
NGC5253; each data point corresponds to a square bin about 10~pc in
size. The colors shown have been subtracted for the underlying galaxy
population and corrected for dust reddening. The curve represents the
locus of models of constant star formation from Leitherer \& Heckman
(1995) and from Bruzual \& Charlot (1995); representative model ages
are marked.
\end{figure}

Extinction--corrected color--color and color--EW(H$\alpha$) plots show that
star formation has been an on-going process for the past $\sim$200~Myr, in the
central $\sim$400~pc of NGC5253 (Figure~2). At present the peak of activity,
identified with the H$\alpha$ peak emission, is located in an area 50--60~pc
in size right north of the dust lane, centered on NGC5253-5 (Figure~1). The
area appears as young as $\sim$5~Myr, as suggested by the large EW(H$\alpha$)
(Calzetti et al.\ 1997), the presence of W-R stars (Schaerer et al.\ 1997), the
scarcity of red supergiants (Campbell \& Terlevich 1984), and the purely
thermal component of the radio emission (Beck et al.\ 1996); observations at
radio wavelengths have revealed the presence of a dusty `supernebula' in the
area, possibly the youngest globular cluster known (Turner, Beck \& Ho
2000). Although this is the most active region in the starburst, with a
SFR$\sim$20~M$_{\odot}$~yr$^{-1}$, it is not the UV-brightest: the area of
peak activity is indeed rather dusty, likely still embedded in the parental
molecular cloud with A$_V\sim$9--35~mag. The bulk of the observed UV emission
is emerging from the less active, relatively unextincted surrounding region,
extending out to $\sim$200~pc in radius. This is the region where star
formation has been going on at a relatively constant pace for
$\sim$100--200~Myr, at the modest rate of 0.2~M$_{\odot}$~yr$^{-1}$. Apart for
some evidence that star formation may be concentrating toward the center, thus
moving inward, there is no other sign of a {\em spatial} evolution of the
starburst in this galaxy.

\begin{figure}[h]
\putplot{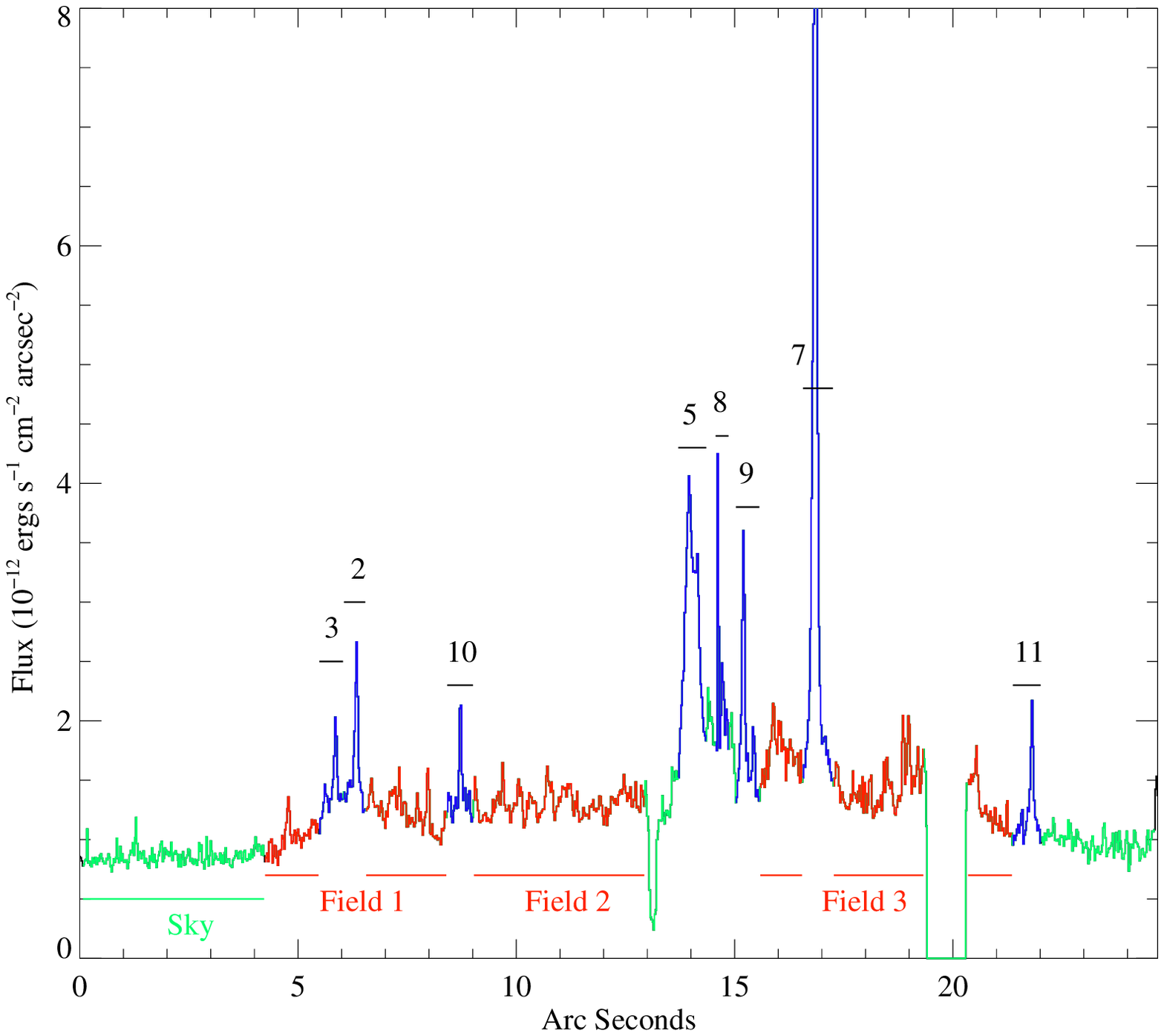}{3.5 cm}{0}{38}{40}{-20}{-100}
\putplot{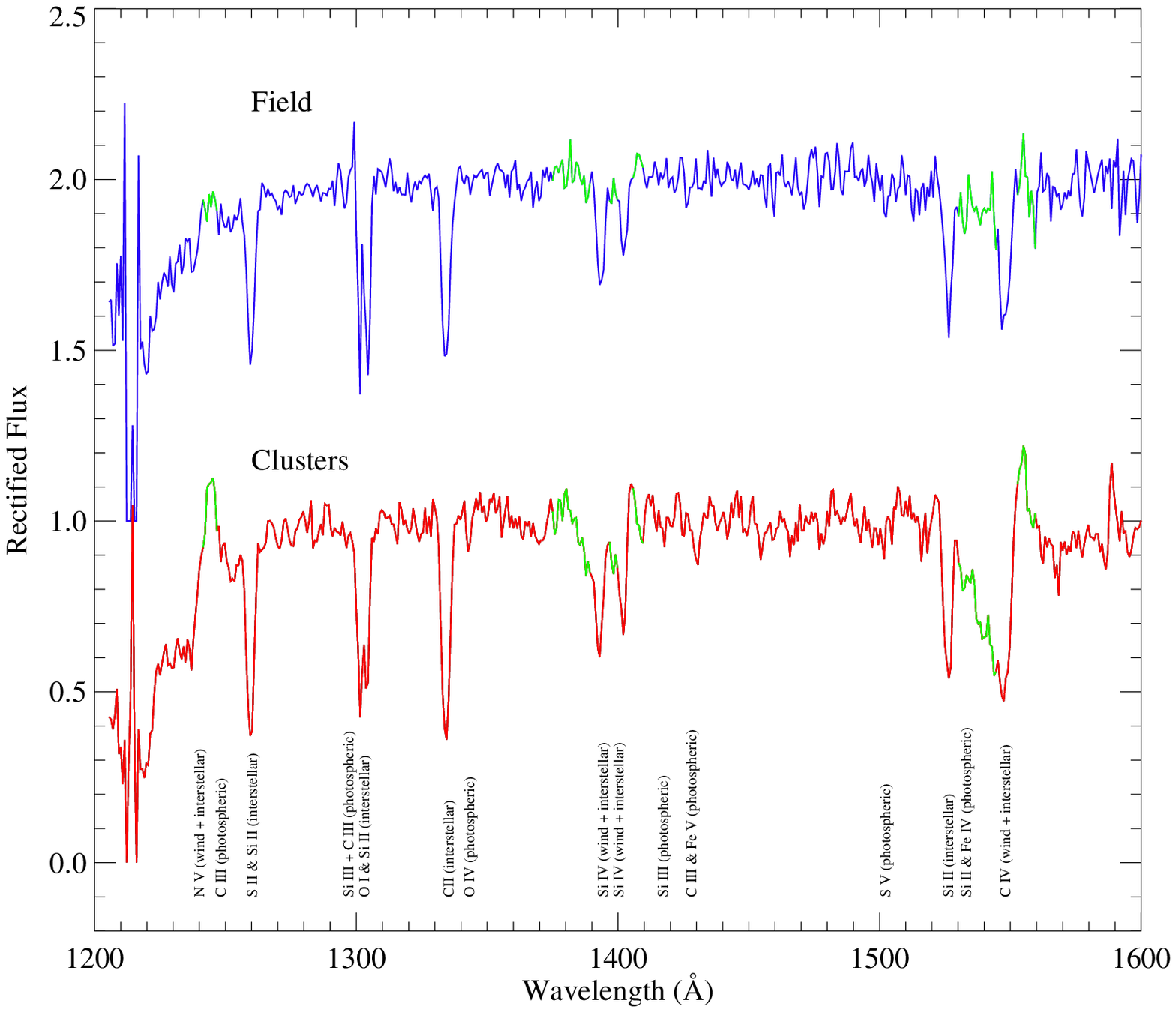}{3.5 cm}{0}{38}{40}{170}{12} 
{\bf Figure 3:} (left) A view of the NGC5253 emission collected by the
STIS long slit, along the spatial direction. The figure shows the
separate components that make up the UV emission of the starburst:
stellar clusters (marked by numbers) and diffuse light (marked as
`Field'), that are well above the background (marked as `Sky').\\
\hspace*{1.4mm} {\bf Figure~4:} (right) The average cluster UV spectrum 
is compared with
the average field UV spectrum.  The two are markedly different; in
particular the spectrum of the field lacks broad--line profiles in
NV(1240~\AA), SiIV(1400~\AA), and CIV(1550~\AA). This difference
excludes scattered cluster light as source of the diffuse UV light
(see the text and Tremonti et al.\ 2001 for a discussion of the nature
of the diffuse light).
\end{figure}

The long-slit of STIS was used to target a number of stellar clusters and the
intracluster UV light in the center of NGC5253, and spectra in the wavelength
range 1150--1700~\AA~ were obtained (Figure~1 and Tremonti et al.\ 2001). Eight
clusters fall within the STIS slit (Figure~3), while the diffuse light (marked
as `Fields' in Figure~3) covers a region of $\sim$220$\times$1.9~pc$^2$ in
total. The eight clusters have ages between 1~Myr and 8~Myr, and masses
between a few 100~M$_{\odot}$ and 4$\times$10$^4$~M$_{\odot}$. The mass of
cluster NGC5253-5 is highly uncertain, depending on uncertain dust
corrections, and could be as high as 6$\times$10$^5$~M$_{\odot}$; possibly, 
the cluster coincides with the `supernebula' detected in the radio by Turner
et al.\ (2000).

The UV spectrum of the field is markedly different from the average spectrum
of the stellar clusters, in that it lacks the broad--line profiles of NV,
SiIV, and CIV, characteristic of O-star winds (Figure~4). The diffuse UV light
makes up between 50\% and 80\% of the total UV emission from local starburst
galaxies (Buat et al.\ 1994; Meurer et al.\ 1995; Maoz et al.\ 1996); 
it has been
suggested to originate from a mode of star formation that is separate from
that of stellar clusters (Meurer et al.\ 1995). Comparison of the STIS UV
spectra of the field with models indicate that, if the field population is
produced by a separate mode of star formation from that of the clusters, this
implies also a different IMF between the two. In particular, while the stellar
clusters are compatible with a Salpeter ($\alpha$=2.35) IMF in the range
1--100~M$_{\odot}$, the field population needs either a steeper-than-Salpeter
IMF slope ($\alpha\sim$3.5) or a standard IMF slope with an upper mass cut-off
at 30~M$_{\odot}$ (Tremonti et al.\ 2001). Steep IMF slopes have been found
also for the field stars of the SMC and LMC, although the values are more
extreme, $\alpha\sim$5, than those derived for NGC5253 (Massey et
al. 1995). An alternative scenario to bimodal star formation is that of
dissolving clusters. The field UV spectrum (Figure~4) is well modelled by a
constant star formation synthetic spectrum from which the young stars'
(younger than 10~Myr) contribution has been subtracted. This `aged' constant
model can easily account for the absence of the broad lines in NV, SiIV, and
CIV, because the O--stars have been removed from it. According to this model,
the lack of massive stars in the field is due to the lack of young stars. The
clusters for which ages have been derived, either from spectroscopy or
photometry, are typically younger than 10--20~Myr. If these are representative
of the clusters' population in the starburst, clusters are generally younger
than the field's stars. Thus, it is possible that {\em all stars form in
clusters}, the clusters dissolve over a timescale of $\sim$10~Myr and their
surviving stars disperse into the field.  Models on the dissolution of compact
stellar clusters in the center of galaxies (Kim, Morris \& Lee 1999) predict
that a 5$\times$10$^3$~M$_{\odot}$ in NGC5253 evaporates in
$\sim$15--20~Myr. This dissolution timescale is in the required ballpark to
make the scenario consistent with the measured ages of NGC5253's clusters.

\section{The Massive Galaxy NGC5236 (M83)}

The starburst in M83 is harbored in a morphologically and dynamically complex
central region (Figure~5). Located within the main bar of the galaxy, the
center of M83 appears to contain a well-defined nuclear subsystem. The bright
optical nucleus is offset from the center of the outer isophotes opposite to
the starburst semi-annulus (Gallais et al.\ 1991), and a nuclear bar may
separate the bright nucleus from the center (Elmegreen, Chromey \& Warren
1998). The possible double nucleus suggests that another galaxy merged with
M83 in the past; the merging event, together with the galaxy's bar, is a
potential trigger of the current central starburst. Dense molecular gas is
concentrated to the north of the starburst semi-annulus (Israel \& Baas 2001),
perhaps a result of material collecting around an inner Lindblad resonance, as
suggested by Gallais et al.\ (1991). This cold material may be feeding the
central starburst (Petitpas \& Wilson 1998).

\begin{figure}[t]
\putplot{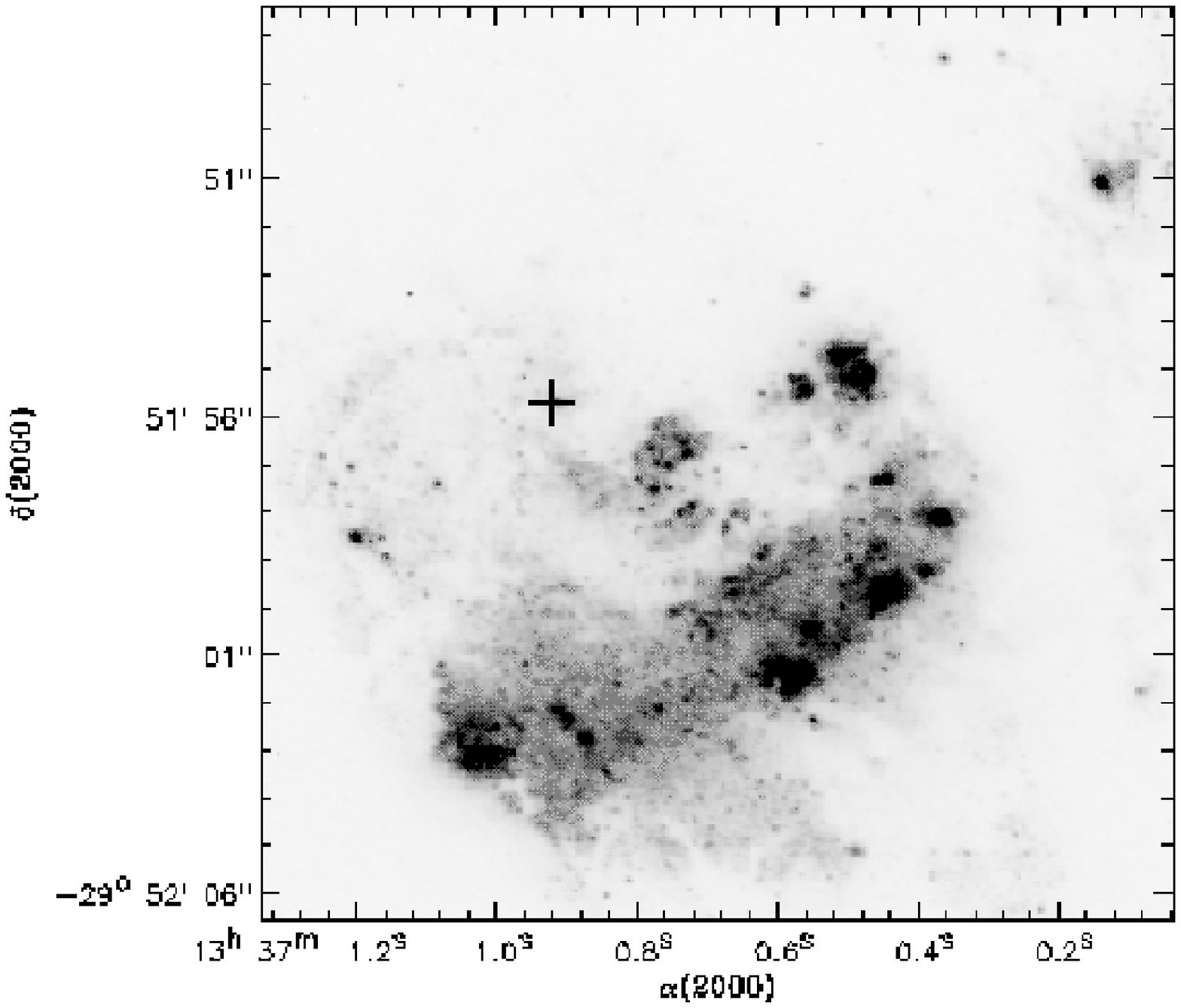}{4.0 cm}{0}{41}{41}{-30}{-150}
\putplot{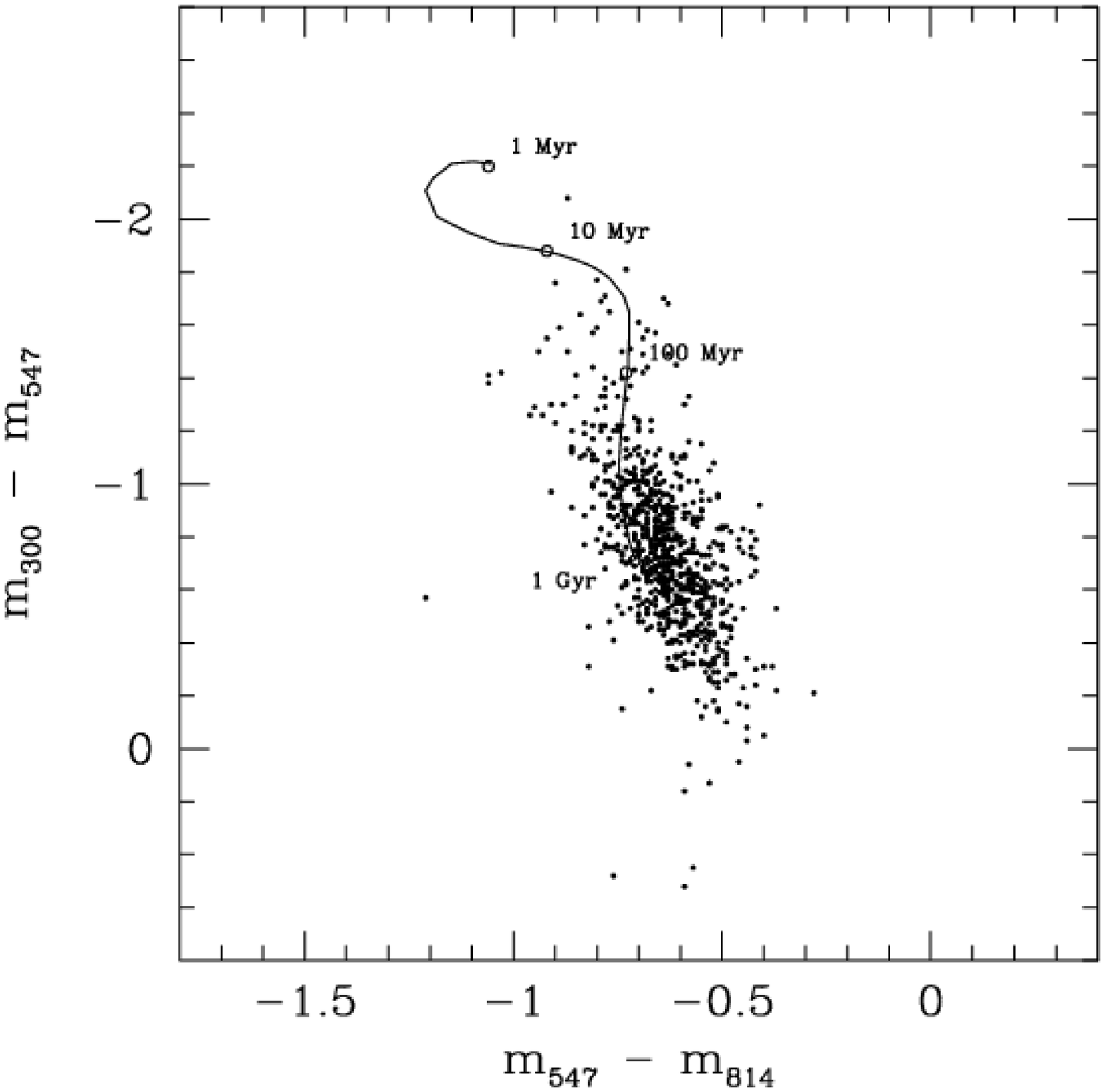}{4.0 cm}{0}{32}{32}{198}{43} 
{\bf Figure 5:}(left) The UV (2900~\AA) WFPC2 image of the central
region in NGC5236, showing the semi-annulus of active star formation
and the position of the optical nucleus (marked with a cross). The
nucleus is very bright at red wavelengths, but faint in the UV. A dark
dust lane is located at the right edge of the image. The image has
size 19$^{\prime\prime}$, which corresponds to a physical size of
$\sim$360~pc. North is up, East is left.\\
\hspace*{1.4mm}  {\bf Figure~6:}(right)
Color--color (UV$-$V versus V$-$I) diagram of the central 300~pc of
M83; each data point corresponds to a square bin about 5~pc in
size. The light from the 45 bright clusters, and from the underlying
galaxy has been subtracted from the images prior to calculating these
colors (that are also corrected for dust reddening). The curve
represents the locus of models of constant star formation from
Leitherer et al.\ (1999) at the appropriate metallicity for the galaxy;
representative model ages are marked.
\end{figure}

The starburst itself, in addition to be of comparable intensity to the
event in NGC5253, is bright at all wavelengths, from the X-ray (Ehle
et al.\ 1998), through the UV (Kinney et al.\ 1993), optical and
near--IR (Gallais et al.\ 1991; Rouan et al.\ 1996), through the
mid--IR (Telesco et al.\ 1993), to the radio (Turner \& Ho
1994). Multiple dust lanes cross the center of this metal--rich
galaxy, although dust is very patchy, as demonstrated by the
UV--brightness of the starburst semi-annulus (Figure~5). Two bright
mid--IR sources, possibly two very young knots of star formation, are
located at the NW edge of the starburst annulus, close to and/or
embedded in a major dust lane (Telesco et al.\ 1993). The UV--bright
region of star formation breaks down into almost 400 clusters brighter
than m$_{UV}$=18~mag (Harris et al.\ 2001), while diffuse light
represents more than 20\% of the total UV emission. The H$\alpha$
morphology closely follows that of the stellar emission, a different
characteristic from NGC5253. 

An analysis of the colors and EW(H$\alpha$) of the UV--bright
population in M83, analogous to that performed for NGC5253 (see
previous section), reveals that star formation has been continuous for
the past $\approx$0.1--1~Gyr (Figure~6). This is not dissimilar to 
what found in NGC5253, although a larger fraction of the population 
in M83 contains stars older than $\sim$500~Myr (Harris et al.\ 2001). 

\begin{figure}[h]
\putplot{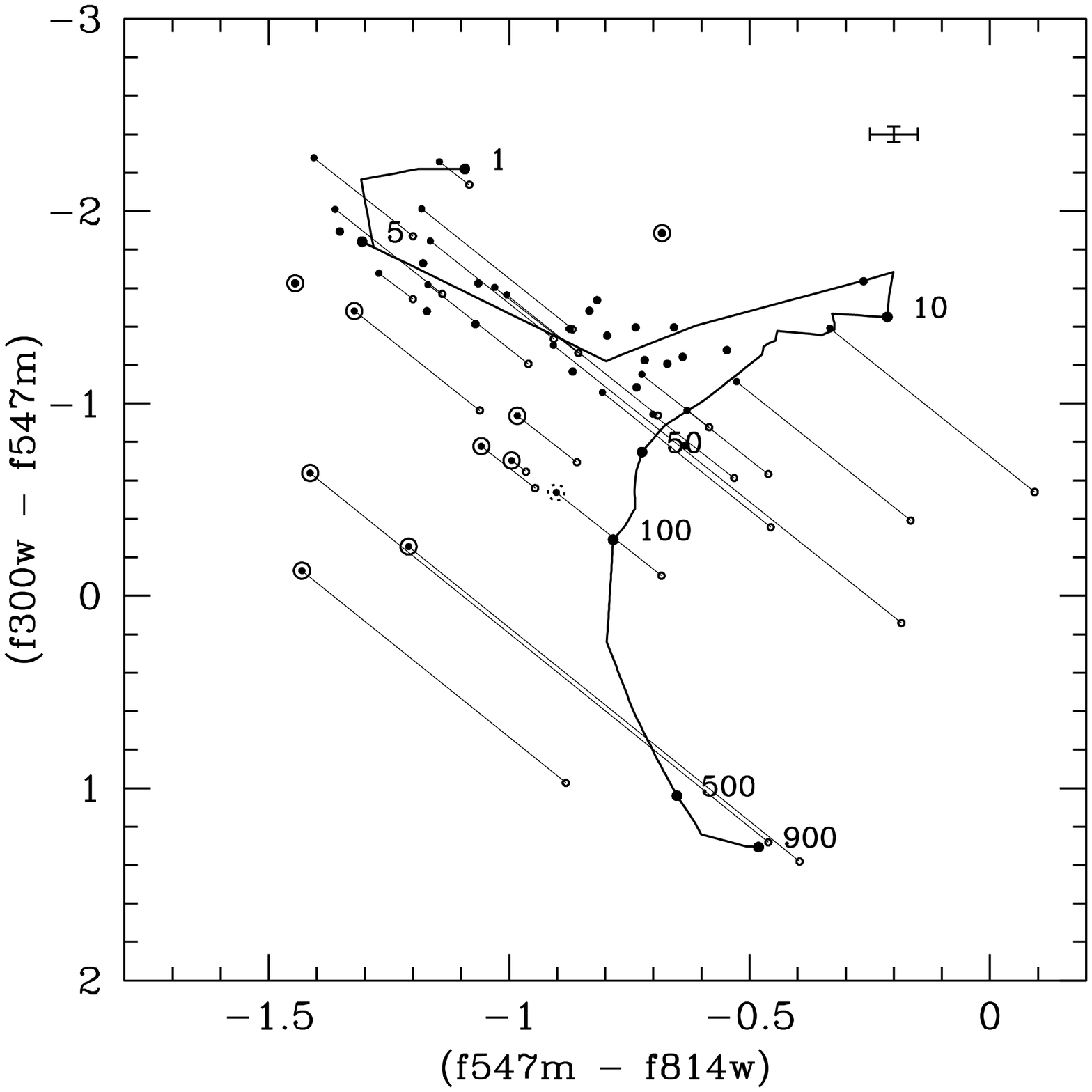}{4.0 cm}{0}{32}{32}{-5}{-137}
\putplot{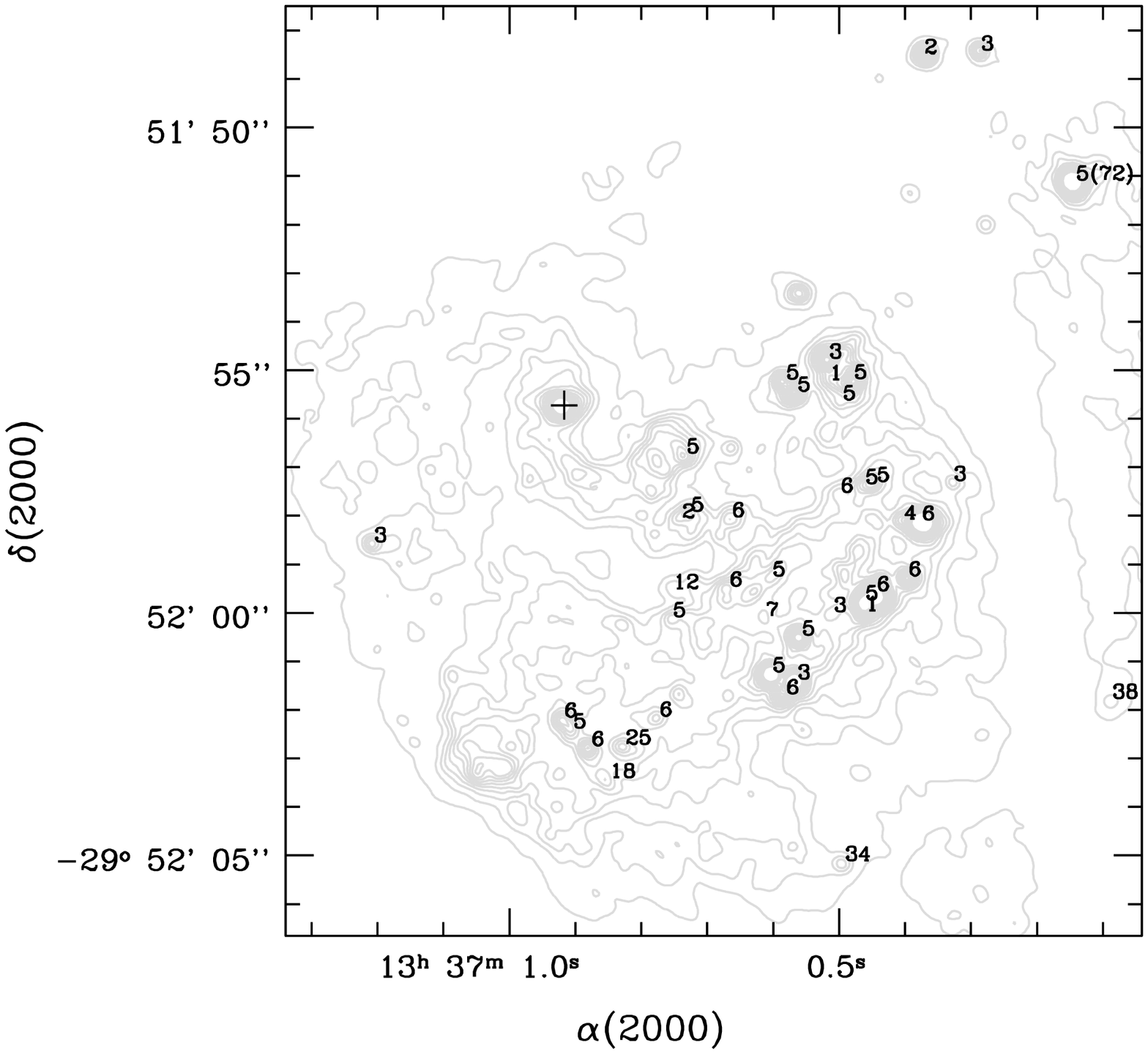}{4.0 cm}{0}{35}{35}{178}{-35} 
{\bf Figure~7:} (left) Color--color (UV$-$V versus V$-$I)
diagram of the 45 brightest clusters in M83. The observed colors are
shown as empty circles; the dust-reddening corrected ones as filled
circles. Lines connect the data points before and after reddening
corrections. Filled circles with no connecting line are clusters
located in areas of low or no reddening. Large empty circles with a
central dot indicate clusters whose age could not be determined from
colors; in these cases the EW(H$\alpha$) was used. The curve
represents the locus of instantaneous burst models from Leitherer et
al. (1999) at the appropriate metallicity for the galaxy;
representative model ages are marked.\\
\hspace*{1.4mm}  {\bf Figure~8:} (right) A map of
the cluster positions in M83; each cluster is marked by its age. The
grey contours trace the V--band image and the position of the bright
optical nucleus is indicated by a cross.
\end{figure}

Among the $\sim$400 clusters detected in the UV with WFPC2, 45 are
bright enough to be detected also in the shallower V and I images,
enabling a study of their age and mass distributions (Harris et
al. 2001). Again, colors and EW(H$\alpha$) were used as complementary
diagnostics for the age of the stellar clusters (Figure~7). The basic
result is that about 75\% of the clusters more massive than
2$\times$10$^4$~M$_{\odot}$ (our low mass completeness limit) are
younger than 10~Myr, and almost 50\% are in the narrow age range
5--7~Myr. No cluster older than $\sim$50~Myr was detected, although
our images are deep enough for the purpose. While the 5--7~Myr old
clusters are distributed across the semi-annulus of star formation,
clusters younger than 5~Myr are preferentially located at the edges of
the annulus (Figure~8).  This suggests that star formation has
propagated from the interior of the annulus to its perimeter. The
5--7~Myr population has possibly evacuated interstellar material from
most of the semi-annulus, and star formation is continuing along the
edges, as also evidenced by the H$\alpha$ morphology.

Along the semi-annulus, there is a trend for older clusters (ages around
10--30~Myr, Figure~8) to be located in the southern-most region, where the
largest of the H$\alpha$ bubbles is also located; whereas the peak of the
current star formation is in the north-west area of the semi-annulus, close to
a major dust lane and to the two mid-IR peaks (Telesco et al.\ 1993). This age
sequence supports earlier suggestions that star formation has been propagating
along the semi-annulus from the south to the north (Gallais et al.\ 1991; 
Puxley et al.\ 1997).

The full mass range for our clusters is
$\sim$10$^3$--8$\times$10$^4$~M$_{\odot}$, and clusters more massive than a
few times 10$^4$~M$_{\odot}$ are not expected to evaporate on timescales
shorter than a few tens of Myr. Thus, the peak in the number of 5--7~Myr old
clusters may suggest a true burst of activity in the recent past of the
galaxy. This is consistent with evolution models already developed for the
center of M83 (Gallais et al.\ 1991, Thatte et al.\ 2000): molecular gas is fed
to the center by the main bar, and once enough mass is accumulated, star
formation occurs. Typical sizes of the clouds would imply burst durations of
10--30~Myr (Efremov \& Elmegreen 1998). Possibly, the current burst is a
transitory flare-up of circum-nuclear star formation, but if gas feeding of
the center by the bar is recurrent or continuous, so would be the bursts of
star formation. The absence of clusters older than 50~Myr in our images would
suggest inter-burst periods $>$50~Myr.

A more accurate understanding of the star formation history in the
center of M83 is currently hampered by our limitations in pinning down
the nature of the diffuse UV light. This is mainly driven by the lack
of UV spectroscopic data. For instance, if the UV spectra were to
indicate that the diffuse population is mainly constituted of stars
older than $\approx$50~Myr, this would reinforce the case for a true
starburst nature of the 5--7~Myr old cluster population. A diffuse
population containing signatures of young O--stars would support a
bimodal star formation scenario (in contrast with NGC5253). Presence
of field stars in the 10--50~Myr age range would re-open the question of
whether the current starburst is a `true burst' or simply part of a
long duty-cycle event, because the scarcity of clusters older than
$\sim$10~Myr could be ascribed to evaporation.

\section{Conclusions}

Preliminary results from the starburst star formation history project
indicate that the morphological differences between the starbursts in
the dwarf NGC5253 and in the massive spiral M83 may reflect
differences in their history as well. In particular, the starburst in
NGC5253 appears to be part of a time-extended ($>$100~Myr) event,
while the starburst in M83 could be a `true' burst of star formation,
a short flare-up that could be separated from the next one by more
than 50~Myr. The latter scenario is still under investigation, and UV
spectroscopy will provide a final answer.

The next steps in the project include analysing the larger sample of 
starbursts, in order to place the above results on a solid statistical 
footing and address the questions posed in the Introduction.

\vspace{0.5cm} D.C. thanks the Scientific Organizing Committee for the
invitation to this stimulating and varied Conference in honor of our
long-time friend and colleague Ken Freeman. This research was funded by
the following grants: HST GO-8232, HST GO-8234, NASA NAG5-9173. Travel
to the conference was funded by the STScI Director's Discretionary
Research Funds.

\end{document}